**Causal inference in multi-cohort studies using the target trial framework to identify and minimize sources of bias**

**Authors:** Marnie Downes, Meredith O'Connor, Craig A Olsson, David Burgner, Sharon Goldfeld, Elizabeth A Spry, George Patton, Margarita Moreno-Betancur

**Affiliations:** Murdoch Children's Research Institute, Melbourne (Downes, O'Connor, Olsson, Burgner, Goldfeld, Spry, Patton, Moreno-Betancur); Department of Paediatrics, The University of Melbourne, Melbourne (O'Connor, Olsson, Burgner, Spry, Patton, Moreno-Betancur); Melbourne Graduate School of Education, University of Melbourne (O'Connor); Centre for Social and Early Emotional Development, School of Psychology, Faculty of Health, Deakin University, Geelong (Olsson, Spry)

**Corresponding author:** Marnie Downes, Clinical Epidemiology and Biostatistics Unit, Murdoch Children's Research Institute, 50 Flemington Road, Parkville, 3052, Victoria, Australia (email: marnie.downes@mcri.edu.au)

**Abstract**

Longitudinal cohort studies, which follow a group of individuals over time, provide the opportunity to examine causal effects of complex exposures on long-term health outcomes. Utilizing data from multiple cohorts has the potential to add further benefit by improving precision of estimates through data pooling and by allowing examination of effect heterogeneity through replication of analyses across cohorts. However, the interpretation of findings can be complicated by biases that may be compounded when pooling data, or, contribute to discrepant findings when analyses are replicated. The "target trial" is a powerful tool for guiding causal inference in single-cohort studies. Here we extend this conceptual framework to address the specific challenges that can arise in the multi-cohort setting. By representing a clear definition of the target estimand, the target trial provides a central point of reference against which biases arising in each cohort and from data pooling can be systematically assessed. Consequently, analyses can be designed to reduce these biases and the resulting findings appropriately interpreted in light of potential remaining biases. We use a case study to demonstrate the framework and its potential to strengthen causal inference in multi-cohort studies through improved analysis design and clarity in the interpretation of findings.

**Abbreviations:** ATP, Australian Temperament Project; ATPG3, Australian Temperament Project, Generation 3; CIS-R, Clinical Interview Schedule – Revised; DAG, directed acyclic graph; DASS-21, Depression Anxiety Stress Scales; GHQ-12, General Health Questionnaire; IPW, inverse probability weighting; RBPCSF, Revised Behavior Problem Checklist Short Form; RCMAS, Revised Children's Manifest Anxiety Scale; RCT, randomized controlled trial; SMFQ, Short Mood and Feelings Questionnaire; STST, Short Temperament Scale for Toddlers; VAHCS, Victorian Adolescent Health Cohort Study; VIHCS, Victorian Intergenerational Health Cohort Study.

Causal inference, understood as the examination of the impact of potential interventions (1), is a common goal in health research, where the ultimate intent is to inform future action that will improve patient or population outcomes. Randomized controlled trials (RCTs) are considered the “gold standard” for causal inference, however, it is often not feasible to implement an RCT design (2). For example, in child and adolescent health, there is a pressing need to identify targets for preventive intervention (3) to counter risk associated with challenges such as obesity (4), mental disorders (5), and allergic diseases (6), that can track forward to adult non-communicable diseases (7-11). However, RCTs are limited in their capacity to provide this evidence, because of the long time-frame for outcomes and ethical issues in randomizing risks like childhood adversity or poverty.

Existing longitudinal cohort studies that follow individuals over long periods of time, offer a viable alternative to address these causal questions (12-14). However, relative to RCTs, observational studies may suffer from higher risks of bias, particularly confounding, which may threaten the validity of causal inferences. For this reason, there is a long history of describing sources of bias that may arise in analyses of cohort studies that aim for causal interpretation (15-19). Building on this foundation, recent methodological advances have developed improved understanding of types of bias as well as methods to tackle them that are now widely used in cohort research (1). The “target trial”, defined as the hypothetical randomized experiment that would answer the causal question of interest, (2) is a powerful tool for guiding the planning, conduct and interpretation of causal analyses in observational data, including single-cohort studies (20-25).

Beyond the potential for causal biases, single-cohort studies may present limitations in terms of sample size, in particular to investigate rare events and exposures (26) as well as effect modification among subgroups. Also, the focus of single cohorts on specific settings, populations and/or epochs may raise questions about the generalizability of findings to other contexts (27, 28). Furthermore, it is possible that estimated effect sizes, even in large well-designed studies, may be exaggerated (12) and not replicable when investigations are repeated in new data (29).

Given their capacity to address these issues, multi-cohort studies – studies involving analyses of individual-level data from multiple independent cohorts – have become increasingly common (30-39), as well as more feasible with platforms arising to support data access (40). There are two key motivations for multi-cohort designs. First, more precise estimation of a causal effect can be achieved by the integration of harmonized data from multiple cohorts into a single dataset, on which analyses are performed directly (*pooled-data analysis, also known as one-step individual participant meta-analysis*) (26, 41), or by synthesis of effect estimates obtained from analyses of individual-level data from each cohort separately (*two-step individual participant meta-analysis*) (42, 43). Second, investigation of effect heterogeneity across contexts, defined here as a difference in true causal effects that are distinct in only one aspect of their definition (e.g., target populations considering different time-periods or places, or outcomes measured at different time-points), may be conducted if this difference is a key distinguishing feature across the cohorts available. This may be achieved by obtaining and comparing estimates in each cohort separately (*replication of analyses with comparison*), by random-effects two-step individual participant meta-analysis with the estimate of the between-study variance providing a measure of heterogeneity, or, when using outcome regression as a confounding-adjustment approach, by pooled-data analysis including a cohort-by-exposure interaction term. Elsewhere (28), we have summarized these approaches and recognised other motivations (and associated methods) for multi-cohort studies such as the capacity to address interrelated components of a complex theoretical model.

It is now timely to draw together these two major shifts in cohort-based health research: the target trial framework and use of multi-cohort designs. As with any causal problem, the reliability of causal inferences arising from the application of any multi-cohort analytic approach is dependent on a thorough understanding of sources of bias. Compared with single-cohort studies, undertaking causal inference in multi-cohort studies faces additional challenges in this regard. When pooling data to improve precision, biases arising in each cohort may be compounded and new biases introduced. Meanwhile, when replicating analyses in each cohort separately to investigate effect heterogeneity, interpretation may be complicated as different biases within each cohort may contribute to and thus distort differences in estimated effects. This makes clear definition of the target estimand and systematic consideration of sources of bias critically important in the conduct of multi-cohort studies. There is currently little guidance on how to do this with the need for thorough planning recently highlighted (44).

The aim of this paper is to propose how the target trial conceptual framework can be extended to address the specific challenges that arise when undertaking causal inference in multi-cohort studies. We review the approach in single-cohort studies, describe the extension to multi-cohort studies, and then use a case study to demonstrate its value in understanding potential biases associated with multi-cohort studies to inform analysis design and interpretation. We conclude by offering some guidance on the selection of existing analytic methods for addressing identified biases.

## The target trial framework in single-cohort studies

The target trial framework involves two steps. The first is to specify the target trial, defined as the hypothetical randomized experiment that would ideally be implemented to answer the research question of interest (2). A detailed description is developed by specifying key protocol components, including eligibility criteria, treatment strategies, assignment procedures, follow-up period, and outcome. Articulation of the target trial yields refined definitions of the research question and corresponding estimand (i.e., the causal effect of interest) (20).

The second step is to consider the assumptions under which one may emulate the target trial with the observational data available to obtain an unbiased estimate of the causal effect. Bias refers to a discrepancy between the estimate of the causal effect of interest from a study (on average over replications) and its true value (which is typically unknown in practice). In emulating each of the target trial protocol components, there are corresponding analysis decisions to consider (analytic sample selection, treatment/exposure measure, selection of confounders, timing of measures, outcome measure), each of which can reduce or introduce bias depending on the assumed causal relationships between the study variables. A causal diagram or directed acyclic graph (DAG) that describes these relationships helps to develop a detailed understanding of potential biases to consider in the target trial emulation (1). Armed with this understanding, analysis decisions can be made to: 1) reduce or counter potential biases; 2) avoid introducing new biases; and 3) allow for thoughtful interpretation of findings considering potentially remaining biases.

Specifically, there are three key causal biases to consider: selection bias, confounding bias, and measurement bias (Table 1). In the DAGs in Figures 1A, 2A and 3A, these biases are represented as arising from biasing paths between exposure and outcome that introduce a non-causal association. While in practice it is generally preferrable to develop a single DAG describing the assumed relationships between all important variables, here we have chosen to present three separate simplified DAGs to demonstrate how DAGs can be used to understand sources of each type of causal bias. We note also that here and in subsequent sections we focus on the simple but common setting of a binary or categorical point treatment (exposure) and a binary or continuous point outcome. We do not consider the more complex settings of sustained treatment strategies and time-to-event outcomes, and related issues like immortal time bias, which are described in-depth elsewhere for single-cohort studies (45) and represent key areas in which the target trial framework is widely known to be helpful in limiting bias risks.

**The target trial framework in multi-cohort studies**

We first consider application of the target trial framework when combining cohorts with the purpose of improving precision using pooled-data analysis, and then highlight how specification of the target trial changes when the framework is applied to investigate effect heterogeneity by replicating analyses in each cohort separately. Although we do not specifically focus on this, it’s worth noting that many of the important considerations when combining data from multiple cohorts are also relevant when combining data from multiple RCTs (46).

When pooling cohort data to improve precision, the initial step, as in single-cohort studies, is to define the causal estimand by specifying the target trial. The next step is to consider emulating the target trial within each cohort separately, where each emulation strategy is tailored specifically to the individual cohort’s design features and observed data available.

Indeed, we propose that the critical point to the application of this framework in the multi-cohort setting is that the specified target trial provides the central point of reference for the identification of potential sources of biases that inform the emulation strategy in each cohort, as opposed to comparing one cohort to another which is the usual tendency. For example, specifying the eligibility criteria characterises the target population for whom inference is sought. This is usually a broad population from which the sample participants of each cohort study have been sourced. Each study sample will only imperfectly represent the target population, and the deviation (and thus potential for selection bias) may differ across cohorts. Recognising this discrepancy is only possible through specification of a single target trial with a clear target population.

More generally, there are two distinct sources of bias when pooling data from multiple cohorts. Firstly, there may be causal biases in the emulation of the target trial in each cohort. We refer to these as "within-cohort" biases. These biases may operate in the same or opposite directions so combining multiple cohorts together can result in a compounding or reduction in bias overall. Secondly, additional biases may be introduced through systematic differences between cohorts in terms of design features such as calendar period, geographic location, recruitment and assessment methods, and personnel. We refer to these as "across-cohort" biases. These biases may take the form of selection bias due to additional common causes of study participation or missing data and the outcome, confounding bias due to additional (measured or unmeasured) common causes of exposure and outcome, and measurement bias due to measurement error being different between the cohorts, which may be exacerbated in the harmonization process required to create a single integrated dataset (see Figures 1B, 2B, 3B respectively).

Examination of within-cohort biases allows evaluation of how well the target trial is emulated in each cohort, while consideration of across-cohort biases enables appraisal of the additional biases arising as a result of data pooling. This dissection of sources of bias would not be possible if cohorts are simply compared to each other, with no clarity about the causal estimand. Using this approach, if it is deemed that the emulation of the target trial is more problematic in one cohort than another, or in other words, if there is considerably more potential for bias in one cohort than another (perhaps due to important study design features that do not meaningfully align or harmonization decisions that require a substantial loss of information), then it may be more appropriate to proceed with a single-cohort study, or in the case of several cohorts, drop one and go ahead with a reduced multi-cohort study. For example, a multi-cohort study using pooled-data analysis to increase precision will, in practice, be appropriate only when the source populations from which the individual cohorts were recruited are relatively well aligned with the target population.

When replicating analyses in each cohort separately to investigate effect heterogeneity, the description of the target trial must explicitly state the specific varying characteristic of interest (e.g., a difference in geographic location of the target population should be included in the eligibility criteria of the target trial). While technically this defines multiple target trials, each one identical except for the specific varying characteristic of interest, for brevity we henceforth continue to refer to a single target trial. Cohort differences with regard to the specific characteristic of interest will not be identified as within-cohort biases, while cohort differences relative to the target trial in all other aspects (e.g., recruitment in different calendar periods) could be identified as within-cohort biases. An estimated difference in the causal effects across cohorts may thus arise due to effect heterogeneity (an actual difference in the true causal effects), different within-cohort biases in each target trial emulation, or random variability. Unfortunately, it is often extremely challenging to pinpoint the source of discrepancies. Use of the target trial framework will help to identify and thus plan to minimise within-cohort biases as much as possible, and outline potential remaining biases to inform interpretation.

**Example case study**

We consider the published study by Spry *et al.* (47), which utilized data from two Australian longitudinal cohort studies (Table 2) to examine the extent to which preconception maternal mental health in adolescence and young adulthood affects children's early life behavioral outcomes. We do not conduct new analyses here. Rather, by articulating a relevant but previously unspecified target trial and considering the emulation strategies implicit in the analytic decisions adopted in the publication, we illustrate how the framework: 1) sheds light on potential within- and across-cohort biases, 2) enables critique of the analysis methods used regarding the extent to which biases were addressed, and 3) illuminates the interpretation of published findings. Although the framework can be used in this way to appraise published studies, it is preferably applied prospectively when planning analyses for a multi-cohort study.

The study aimed to obtain precise estimates of the causal effects of preconception maternal mental health problems in both adolescence and young adulthood on infant emotional reactivity at 1 year postpartum, through a pooled-data analysis of the two cohorts. Estimates were obtained from a logistic regression model adjusted for cohort and selected confounders, fitted within a generalised estimating equation framework to account for within-family clustering. These estimates can be interpreted as conditional causal odds ratios under the assumption of no effect modification by any of the adjusting variables and that the outcome regression model is correctly specified (1), in addition to the usual causal assumptions of conditional exchangeability, consistency, and positivity, including no bias due to measurement error or missing data. Results of the pooled-data analysis were reported as the primary finding, and as is good practice, replication of analyses were reported as secondary analyses to explore the consistency of findings across cohorts. As expected, pooled-data analysis achieved superior precision and some degree of discrepancy between the cohort-specific causal effect estimates was observed (Table 3).

**Application of the target trial framework to the case study**

Table 4 outlines a proposed target trial and corresponding emulation strategies for VIHCS and ATPG3 implicit in the analysis approach described in Spry *et al.* (47). Considering each protocol component in turn, we first identify the within-cohort biases (e.g., Figures 1A, 2A, 3A), which may have been compounded in the pooled-data analysis, and may explain observed discrepancies in the replication of analyses. We then describe additional across-cohort biases that may have arisen in the pooled-data analysis (e.g., Figures 1B, 2B, 3B). In these DAGs, an additional node representing a cohort indicator is introduced as a proxy for all those cohort characteristics, whether measured or unmeasured, that are fixed within a cohort but vary across cohorts.

A second case study (48) is presented in the Supplementary Material, to highlight additional considerations in the application of the framework when aiming to investigate effect heterogeneity using replication of analyses.

*Eligibility criteria*

The target population of interest is defined as young adolescent females (13 years) in Victoria, Australia in the 1990s. The original cohorts of VAHCS and ATP were designed to recruit close-to-representative samples of this target population, however, individuals from families who could not speak English were excluded, hence there is potential for within-cohort type 2 selection bias (see Table 1) if the effect of the exposure on the outcome in mothers from non-English speaking families is different to the effect in those from English speaking families.

There is further potential for within-cohort selection bias, not only because some eligible individuals might have not consented to participate in the original cohorts, but also because those who did may still not have been eligible or consented to participate in the subsequent intergenerational studies. Specifically, eligibility for recruitment to VIHCS/ATPG3 required female participants of VAHCS/ATP to have fallen pregnant and had a live birth during the screening period. There is potential for type 1 selection bias (see Table 1) if falling pregnant and having a live birth outside the age range of 29–35 years is a common effect of the exposure and unmeasured variables (e.g., problematic maternal drug use during adolescence and young adulthood) that are also causes of exposure and outcome. Note that in settings like this where the outcome is defined after a post-treatment index event (here, live birth), the choice of estimand is a complex problem often without a completely satisfactory solution (49). The target trial specified here examines so-called “total effects” but whether this is the most meaningful estimand is debatable. Furthermore, eligible individuals might have not consented to participate in VIHCS/ATPG3. The DAG in Figure 1A depicts one possible example of type 1 selection bias due to non-participation, with maternal professional status in young adulthood a common cause of VIHCS/APTG3 study participation and the outcome.

Conditioning on study participation leads to a biasing path between the exposure and the outcome via maternal professional status.

Additionally, given the intergenerational nature of these studies, and despite high retention rates, there is potential for within-cohort type 1 selection bias due to missing data in any analysis variable, for example due to study dropout, if the analysis is restricted to participants without missing data (i.e., “complete cases”). This could be depicted in a DAG similar to Figure 1A with the study participation node replaced with a “complete case” indicator. Spry *et al*. (47) used multiple imputation to handle missing data (50), which can address potential selection bias if auxiliary variables that predict both missingness and dropout are included.

There is also potential for across-cohort type 1 selection biases due to systematic between-cohort differences in factors that predict both VIHCS/ATPG3 study participation and the outcome. For example, the two original cohorts utilized different recruitment strategies: VAHCS in adolescence through secondary schools and ATP in infancy through maternal child health services. Different factors during these distinct life stages may have influenced family decisions to consent to participate in VAHCS/ATP, which was a prerequisite to participating in VIHCS/ATPG3. Furthermore, other between-cohort differences such as period of recruitment may be predictors of the outcome. Together, this creates a further biasing path between exposure and outcome via cohort that is introduced by the restriction to VIHCS/ATPG3 participants (Figure 1B). In supplementary analyses of complete cases, cohort differences in factors that are common causes of missing data and the outcome, such as study personnel and assessment tools/methods, may have introduced additional type 1 selection bias (again similar to the DAG in Figure 1B with the VIHCS/ATPG3 study participation node replaced with a “complete case” indicator).

*Treatment strategies*

The exposure of interest is preconception maternal mental health problems in adolescence (13–18 years) and young adulthood (19–29 years), considered here for simplicity as a categorical point-exposure. When considering the target trial, this is an example of an imprecisely-defined intervention. While an important intervention target, it is not clear how mental health problems might be changed or treated. It could be medication, individual therapy and/or population-level interventions. This lack of precise articulation within the target trial complicates the interpretation of findings and the selection of confounders (51). While a common issue in studies asking complex causal questions of this nature, it does not preclude the examination of such questions, rather it reinforces the need for clear causal thinking, and in particular, the benefits of using the target trial framework to highlight study strengths and limitations (20).

Beyond the issue of imprecisely-defined interventions, measuring mental health problems is inherently challenging. It is not directly observable nor can any informant provide the 'true' perspective (52). Symptoms can also manifest differently over contexts and time (53). While well-regarded assessment tools are available, the best such tools can hope to achieve is to approximate the intended underlying construct, with some degree of measurement error unavoidable. In this example, different measurement tools were used across cohorts and also across waves within each cohort, and it is possible that these capture overlapping but slightly different constructs. Additionally, both cohorts relied on mother's self-report only, which while valuable, may be influenced by feelings of guilt, shame or embarrassment (54). Such measurement issues in the exposure could lead to within- and across-cohort measurement biases as is described in more detail below for the outcome measure.

*Assignment procedures*

In the target trial, individuals would be randomized to experiencing mental health problems or not. Clearly, however, even if a specific intervention had been defined to achieve this, it would be unethical to randomize it. Focusing on the available data, a number of measured confounders were identified, based on prior evidence in the literature, representing socioeconomic circumstances and adolescent smoking (Figure 2A). Even after adjusting for these confounders, there remains potential for residual within-cohort confounding bias due to biasing paths via unmeasured confounders (also Figure 2A). There is also potential for residual within-cohort confounding bias due to confounder measurement error as a result of using proxies (e.g. high-school completion and divorce used as proxies for socioeconomic circumstances), and inaccurate reporting of sensitive information (e.g., smoking history or family divorce).

There is potential for exacerbation of residual within-cohort confounding bias due to measurement error in the necessary harmonization of confounder variables. For example, measurement of maternal education was not aligned between VIHCS and ATPG3 hence there was some loss of information in the blunt harmonized measure of “never completed” versus “ever completed” high school. Finally, some additional confounders may have been available in only one of the two cohorts so were not included in the adjustment set, thus increasing the potential for confounding bias.

Additional across-cohort confounding biases may arise in the pooled-data analysis due to additional biasing pathways between exposure and outcome because of systematic between-cohort differences in measured or unmeasured design aspects (Figure 2B). For example, recruitment of mothers and their infants took place in 2006–2013 for VIHCS and in 2012–2018 for ATPG3, with period effects on exposure and outcome assessment potentially inducing additional confounding bias.

*Follow-up period*

Consistent with the target trial, follow-up in both cohorts commenced when mothers were in adolescence and concluded at 1 year postpartum. There is potential for within-cohort measurement error (and thus bias) as a result of the exposure and outcome measurements not being taken at the exact ages specified in the target trial for all participants.

Additional across-cohort measurement bias is possible due to systematic between-cohort differences in the timing of exposure and outcome measurements. For example, there was variation between (as well as within) cohorts in the timing of measurement of the 1 year postpartum outcome.

*Outcome*

Both cohorts used the Short Temperament Scale for Toddlers (STST) via maternal report at 1 year postpartum with a mean score of ≥4 indicative of heightened offspring infant emotional reactivity, in contrast to the target trial, which would triangulate perspectives from parents and clinicians with direct observation of infant behavior. Therefore, there is potential for measurement error and thus bias in the use of STST as a proxy for a more comprehensive assessment of infant emotional reactivity. Additionally, Spry *et al*. (47) identified that maternal report of infant outcomes may be affected by a mother's mental state such that depressed mothers perceive their infant as more reactive. This measurement bias due to inaccurate reporting is depicted in Figure 3A as a biasing path between the exposure and the outcome via the measured outcome.

There could be additional across-cohort bias in the pooled-data analysis if there were systematic between-cohort differences in measurement of the outcome (Figure 3B). This is less of a concern here, however, since infant emotional reactivity was measured consistently in both cohorts. If different outcome measures were used across cohorts, the DAGS in Figures 3A and 3B would become more complex, depicting the cohort-specific outcome measures, $Y_i^*$ $(i = 1, \ldots, k)$ and the pooled outcome measure $Y^{**}$, respectively.

**Guidance for planning and reporting analyses to address identified biases**

Once within- and across-cohort biases have been identified via application of the target trial framework, a comprehensive analysis plan to counter them can be developed (55). It will not be possible to completely eliminate all causal biases, but the aim is to devise an analysis strategy that diminishes bias threats as much as possible and, importantly, does not introduce new ones, for example, through conditioning on a “collider” (see Table 1). This section provides general guidance on the selection of existing analytic approaches for addressing potential sources of each type of causal bias in multi-cohort studies.

*Selection bias*

Missing data methods such as multiple imputation (50, 56) and inverse probability weighting (IPW) (57, 58) are widely utilized for countering selection bias due to study participation, loss to follow-up and other missing data. Modern implementations of multiple imputation (59, 60) provide a flexible approach to handle complex multivariable missingness problems, allowing specification of tailored imputation models including all analysis variables in addition to predictors of incomplete variables that may also be predictors of missingness (e.g., maternal professional status in Figure 1A). In pooled analyses, the cohort indicator is one such variable (Figure 1B), and as such should be included as a covariate in multiple imputation. Alternatively, in replication of analyses, it is recommended that multiple imputation be performed in each cohort separately (61), particularly when cohorts represent distinct populations, settings and/or time periods. This was the approach taken by Spry *et al.* (47) and allows imputation models to be tailored to each cohort.

It is worth noting that there is no preferred method for addressing selection bias due to multivariable missing data. The most appropriate approach for a given problem depends primarily on the target estimand and the multivariable missingness assumptions which determine identifiability. Recent work has used “missingness DAGs” (DAGs expanded to include variable-specific missingness indicators) to depict missingness assumptions in a single-cohort setting and then derive the identifiability of a range of estimands based on these assumptions (62, 63). These results can in turn be used to guide the method for handling missing data (62-64). Extensions to the multi-cohort setting would be valuable.

Meta-analysis methods for transporting inferences from multiple RCTs to a new target population (65-67) may be useful to specifically address type 2 selection bias in pooled-data analysis. However, extensions are needed to allow simultaneous adjustment for confounding and measurement bias (65, 66, 68), both of which are important considerations when working with observational data.

*Confounding bias*

There are two classes of analytic approaches for addressing confounding bias: conditioning-based methods (e.g., multivariable outcome regression) and standardisation-based methods (or “G-methods”, e.g., IPW, g-computation) (1). These methods require modelling either the outcome or the exposure based on the selected confounder set. So-called “doubly-robust” methods use models for both processes and reduce the risk of model misspecification bias due to their good performance when at least one of the models is consistently estimated and because they can be coupled with machine learning (69). Regardless of the analytic method, it is critical to include the cohort indicator as an additional confounder in a pooled-data analysis to address any across-cohort confounding bias due to systematic differences in study design features. This was the approach taken by Spry *et al.* (47), effectively closing the additional non-causal path between exposure and outcome via cohort that is opened by combining the two cohorts (Figure 2B).

*Measurement bias*

Addressing sources of measurement bias is complex, particularly in multi-cohort studies where harmonization of measures to create a single integrated dataset is required and may entail simplification of measures (e.g., collapsing categories) to find a minimum common ground among cohorts. Specialised methods to handle meansurement error such as “regression calibration” (70) exist, though these usually require strong assumptions and validation samples. As with confounding bias, inclusion of the cohort indicator in a pooled-data analysis will address across-cohort measurement biases due to systematic between-cohort differences by closing the additional non-causal path between exposure and outcome via cohort (Figure 3B).

*Sensitivity analyses*

Sensitivity analyses play an important role in exploring the robustness of findings to assumptions underpinning the chosen analytic approach. This is particularly relevant for multi-cohort studies where it may be useful to explore the ramifications of data integration, by repeating the analysis under different harmonization decisions. Formal quantitative bias analysis approaches (71, 72) that quantify the direction and magnitude of systematic biases may also be valuable. These methods are increasingly recommended for individual cohort analyses and we recommend they also be applied to pooled-data analyses, with their application guided by DAGs expanded with the cohort indicator (e.g., Figures 1B, 2B, 3B).

**Conclusion**

The target trial framework can be seen as a powerful tool for improving the conduct of causal inference in observational studies through requiring explicit and detailed definition of the target estimand and facilitating systematic assessment of potential sources of bias. From this perspective, we have described the application and utility of this framework in multi-cohort studies, clarifying that the target trial is the central reference point for identifying biases as opposed to comparing studies to each other. Specifically, this approach enables identification of biases within each cohort individually and those that may be introduced when combining data from multiple cohorts. Disentangling biases arising from different sources helps to better harness the risk of bias in analyses and informs the interpretation of findings, in particular discrepant findings across cohorts. As such, use of the target trial framework in multi-cohort studies as described here can help strengthen causal inferences through improved analysis design, transparency in the assumptions and clarity in reporting and interpretation. We expect continued refinement of this framework as we learn from its application to a wider range of multi-cohort studies in the future.

## Tables and figures

**Table 1:** A summary of the three key causal biases that are important to consider when working through a target trial emulation: selection bias, confounding bias, and measurement bias.

| Bias | Description | Target trial protocol component where emulation can give rise to bias (related analytic design aspect) |
|---|---|---|
| *Selection bias* | Occurs when the sample used for analysis (the analytic sample) is not representative of the population for whom inference is sought (the target population) due to, for example, individuals with certain characteristics being more likely to not participate or be lost from the study over time. Bias can arise in the selection of the cohort sample (the sample of participants recruited into the cohort, i.e., those who are both eligible and consent to participate) from the target population, and also in the selection of the analytic sample from the cohort sample.<br><br>Lu *et al.*(68) distinguishes between two types of selection bias:<br>• Type 1 selection bias arises due to restricting to or conditioning on one (or more) level(s) of a common effect of two variables (known as a "collider"), one of which is either the exposure or a cause of the exposure, and the other is either the outcome or a cause of the outcome. Formally, in a causal diagram or directed acyclic graph (DAG), type 1 selection bias is represented as arising from a non-causal (biasing) path that becomes open after restricting or conditioning on a collider (see Figure 1A).<br>• Type 2 selection bias arises due to restricting to one (or more) level(s) of an effect measure modifier (a variable by which the magnitude of the effect of the exposure on the outcome differs across strata). It is not possible to depict by way of biasing paths in a DAG as for the other biases. | *Eligibility criteria* (analytic sample selection) |
| *Confounding bias* | Arises from differences between the exposure groups in terms of individual, pre-exposure characteristics that are also related to the outcome. Formally, in a DAG, confounding bias results from an open non-causal (biasing) "backdoor" path between the exposure and the outcome that remains even if all arrows pointing from the exposure to other variables (the descendents of the exposure) are removed (i.e., the path has an arrow pointing into the exposure) (1) (see Figure 2A for example). | *Assignment procedures* (confounder selection) |
| *Measurement bias* | Refers to bias that arises as a result of measurement error, which reflects a discrepancy between the measured value of a quantity and its true value as determined by a gold standard instrument. This includes misclassification in the case of categorical data. Measurement bias can be depicted in a DAG in many ways; Figure 3A provides an example of bias arising from measurement error in the outcome, with a non-causal (biasing) path that becomes open when using the measured outcome. | *Treatment strategies* (exposure measurement)<br>*Follow-up period* (timing of measurements)<br>*Outcome* (outcome measurement) |

**Table 2:** A summary of the two Australian longitudinal cohort studies utilized in the published multi-cohort study by Spry *et al.* (47) to examine the extent to which preconception maternal mental health in adolescence and young adulthood affects children's early life behavioral outcomes.

| | **The Victorian Intergenerational Health Cohort Study (VIHCS)** | **The Australian Temperament Project, Generation 3 (ATPG3)** |
|---|---|---|
| Type of study | A prospective longitudinal study (73) arising from an existing cohort study, The Victorian Adolescent Health Cohort Study (74). | A prospective longitudinal study (75) arising from an existing cohort study, The Australian Temperament Project (76). |
| *Original cohort study:* | | |
| Recruitment | VAHCS commenced in 1992 and recruited a sample of mid-secondary school students aged 14–15 years (*N*=1943; 1000 female) from Victoria, Australia. | ATP commenced in 1983 and recruited a sample of infants aged 4–8 months (*N*=2443; 1170 female) along with their parents, through maternal and child health centres in urban and rural local government areas in Victoria, Australia. |
| Follow-up | Participants were assessed six-monthly during adolescence and three times in young adulthood. | Families were invited to complete mail surveys every 1–2 years until 19–20 years of age and every 4 years thereafter. |
| *Subsequent intergenerational cohort study:* | | |
| Screening and recruitment | Six-monthly screening of VAHCS female participants between 2006 and 2013 (age 29–35 years) for pregnancies or recently born infants. | Six-monthly screening of ATP female participants between 2012 and 2018 (age 29–35 years) for pregnancies or recently born infants. |
| Follow-up | VIHCS participants were asked to complete telephone interviews in trimester 3, 2 months and 1 year postpartum for each infant born during screening, with follow-up assessments continuing into offspring childhood and adolescence. | ATPG3 participants were asked to complete interviews in trimester 3, 2 months and 1 year postpartum for each infant born during screening, with follow-up assessments continuing into offspring childhood. |

(VIHCS: Victorian Intergenerational Health Cohort Study; ATPG3: Australian Temperament Project Generation 3; VAHCS: Victorian Adolescent Health Cohort Study; ATP: Australian Temperament Project).

**Table 3:** Estimated causal effects (expressed as odds ratios with 95% confidence intervals) of preconception maternal mental health problems in adolescence and young adulthood on heightened offspring infant emotional reactivity in pooled-data analysis and replication of analyses reported in Spry *et al.* (47). Analyses included adjustment for the identified confounders listed in Table 4 and Figure 2.

| | **Odds ratio** | **95% CI** |
|---|---|---|
| *Pooled-data analysis* | | |
| No preconception maternal mental health problems (unexposed) | | |
| Preconception maternal mental health problems (exposed) | | |
| In adolescence only | 1.3 | (0.9, 2.0) |
| In young adulthood only | 1.3 | (0.7, 2.1) |
| In adolescence and young adulthood | 2.1 | (1.4, 3.1) |
| *Replication of analyses* | | |
| VIHCS | | |
| No preconception maternal mental health problems (unexposed) | | |
| Preconception maternal mental health problems (exposed) | | |
| In adolescence only | 1.2 | (0.6, 2.3) |
| In young adulthood only | 1.8 | (0.8, 3.9) |
| In adolescence and young adulthood | 2.4 | (1.3, 4.2) |
| ATPG3 | | |
| No preconception maternal mental health problems (unexposed) | | |
| Preconception maternal mental health problems (exposed) | | |
| In adolescence only | 1.5 | (0.8, 2.7) |
| In young adulthood only | 0.9 | (0.4, 2.0) |
| In adolescence and young adulthood | 1.9 | (1.1, 3.4) |

**Table 4:** Proposed target trial and emulation strategies implicit in the statistical analysis approach of Spry *et al.* (47), for precise estimation of the causal effects of preconception maternal mental health problems in both adolescence and young adulthood on offspring infant emotional reactivity.

| | | Emulation strategies | |
|---|---|---|---|
| **Protocol component** | **Target trial** | **VIHCS** | **ATPG3** |
| ***A. Eligibility criteria*** | **Target population:**<br>Young adolescent females (13 years of age) in Victoria, Australia in the 1990s | **Analytic sample selection:**<br>Female VAHCS study participants (a sample of 1992 Victorian mid-secondary school students aged 14–15 years), who subsequently reported pregnancy or recently born infant between 2006 and 2013 (29–35 years) when screened, and consented to participate in VIHCS.<br>**Approach to handling missing data and other potential sources of selection bias:**<br>All VIHCS participants were retained in the sample regardless of missing data via use of multiple imputation.There was no attempt to address selection bias due to conditioning on VAHCS participation, having a live birth and VIHCS participation. | **Analytic sample selection:**<br>Female ATP study participants (recruited through rural and urban Victorian Maternal and Child Health centres at 4–8 months of age in 1983), who subsequently reported pregnancy or recently born infant between 2012 and 2018 (29–35 years) when screened, and consented to participate in APTG3.<br>**Approach to handling missing data and other potential sources of selection bias:**<br>All ATPG3 participants were retained in the sample regardless of missing data via use of multiple imputation. There was no attempt to address selection bias due to conditioning on ATP participation, having a live birth and ATPG3 participation. |
| ***B. Treatment strategies*** | **Treatment arms in the trial:**<br><u>Intervention arms:</u><br>Preconception maternal mental health problems in:<br>1. Adolescence (13–18 years) only<br>2. Young adulthood (19–29 years) only<br>3. Adolescence (13–18 years) and young adulthood (19–29 years)<br><u>Comparator arm:</u><br>No preconception maternal mental health problems in | **Treatment/Exposure measure :**<br><u>Intervention arms:</u><br>The presence of any mental health problems at ≥1 wave in:<br>1. Adolescence (14–18 years; VAHCS waves 2–6) only<br>2. Young adulthood (19–29 years; VAHCS waves 7–9) only<br>3. Adolescence (VAHCS waves 2–6) and young adulthood (VAHCS waves 7–9)<br><u>Comparator arm:</u><br>No mental health problems at any wave in adolescence or young adulthood (VAHCS waves 2–9) | **Treatment/Exposure measure:**<br><u>Intervention arm:</u><br>The presence of any mental health problems at ≥1 wave in:<br>1. Adolescence (13–18 years; ATP waves 10–12) only<br>2. Young adulthood (19–28 years; ATP waves 13–15) only<br>3. Adolescence (ATP waves 10–12) and young adulthood (ATP waves 13–15)<br><u>Comparator arm:</u><br>No mental health problems at any wave in adolescence or young adulthood (ATP waves 10–15) |
| | adolescence (13–18 years) or young adulthood (19–29 years) | Mental health problems measure:<br>Waves 2–7: CIS-R ≥12<br>Waves 8–9: GHQ-12 ≥3 | Mental health problems measure:<br>Wave 10: SMFQ ≥11 or RBPCSF mean ≥1<br>Waves 11–12: SMFQ ≥11 or RCMAS mean ≥1<br>Waves 13–15: DASS-21, Depression ≥7 or Anxiety ≥6 or Stress ≥10 |
| ***C. Assignment procedures*** | **Randomisation strategy:**<br>Randomisation at recruitment<br>without blind assignment | **Selection of confounders:**<br>Confounder (harmonized self-reported measure)<br>• Mother's parent's high school completion (neither parent vs. at least one parent completed)<br>• Mother's parent's divorce/separation during or before adolescence (ever vs. never divorced/separated)<br>• Mother's high school completion (ever vs. never completed)<br>• Mother's adolescent smoking (daily smoking at ≥1 adolescent wave vs. no daily smoking)<br>• Mother's history of divorce/separation (ever vs. never divorced/separated)<br><br>**Approach to confounding adjustment:**<br>Oucome regression | **Selection of confounders:**<br>Confounder (harmonized self-reported measure)<br>• Mother's parent's high school completion (neither parent vs. at least one parent completed)<br>• Mother's parent's divorce/separation during or before adolescence (ever vs. never divorced/separated)<br>• Mother's high school completion (ever vs. never completed)<br>• Mother's adolescent smoking (daily smoking at ≥1 adolescent wave vs. no daily smoking)<br>• Mother's history of divorce/separation (ever vs. never divorced/separated)<br><br>**Approach to confounding adjustment:**<br>Outcome regression |
| ***D. Follow-up period*** | **Start and end times:**<br>Start: At randomisation (mother aged 13 years)<br>End: Child aged 1 year | **Timing of measures:**<br>Start: VAHCS wave 2 (mother aged 14–15 years old)<br>End: VIHCS wave 3 (child aged 1 year old) | **Timing of measures:**<br>Start: ATP wave 10 (mother aged 13–14 years old)<br>End: ATPG3 wave 3 (child aged 1 year old) |
| ***E. Outcome*** | **Outcome measure:**<br>Heightened offspring infant emotional reactivity determined through triangulation of parent-report, clinician ratings and direct observation of infant behavior | **Outcome measure:**<br>STST via maternal report at 1 year postpartum, mean score ≥4 | **Outcome measure:**<br>STST via maternal report at 1 year postpartum, mean score ≥4 |
| ***F. Causal contrasts of interest and*** | Conditional odds ratio comparing risk of heightened offspring infant | | |

| | | |
|---|---|---|
| | ***causal effect measure*** | emotional reactivity in each intervention arm relative to the comparator arm in the target population |

(VIHCS: Victorian Intergenerational Health Cohort Study; ATPG3: Australian Temperament Project Generation 3; VAHCS: Victorian Adolescent Health Cohort Study; ATP: Australian Temperament Project; CIS-R: Clinical Interview Schedule – Revised; GHQ-12: General Health Questionnaire; SMFQ: Short Mood and Feelings Questionnaire; RBPCSF: Revised Behavior Problem Checklist Short Form; RCMAS: Revised Children's Manifest Anxiety Scale; DASS-21: Depression Anxiety Stress Scales; STST: Short Temperament Scale for Toddlers).

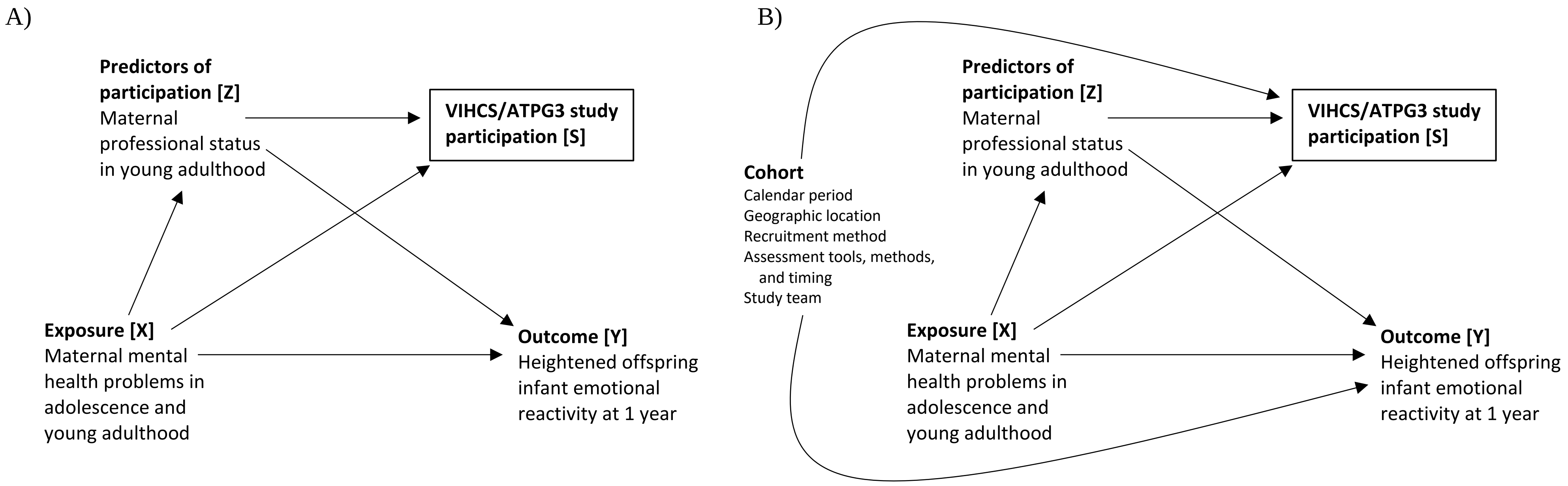

**Figure 1:** Directed Acyclic Graphs (DAGs) depicting examples of: A) type 1 selection bias in a single cohort (within-cohort selection bias); and B) additional type 1 selection biases in pooled-data analyses of multiple cohorts (across-cohort selection bias).

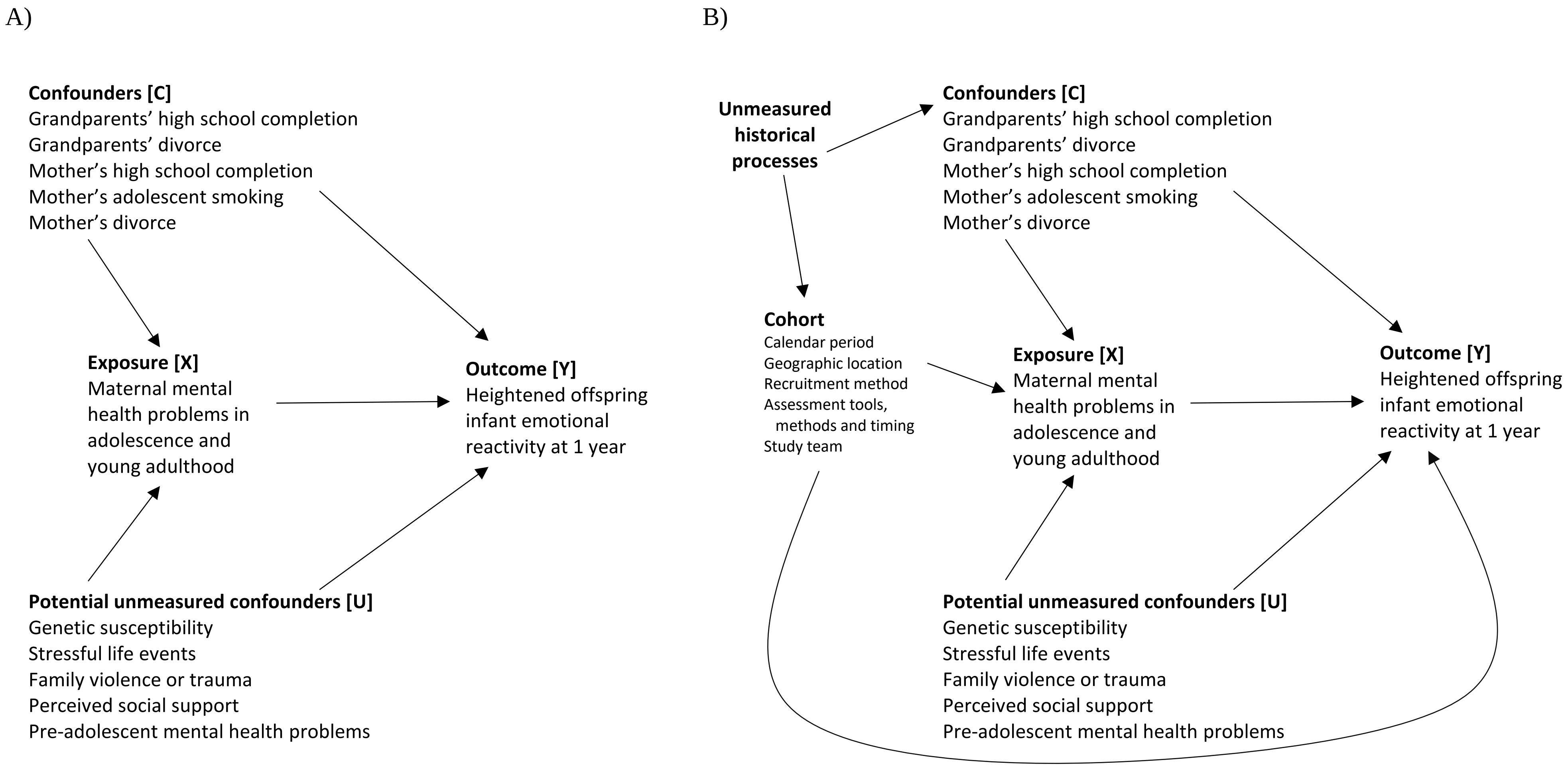

**Figure 2:** Directed Acyclic Graphs (DAGs) depicting examples of: A) confounding bias in a single cohort (within-cohort confounding bias); and B) additional confounding biases in pooled-data analyses of multiple cohorts (across-cohort confounding bias). (Note: following the work of VanderWeele & Robinson (51) and Moreno-Betancur *et al.* (77), we include a node in B) to represent the unmeasured historical processes that are common causes of the cohort characteristics and the individual characteristics considered as confounders.)

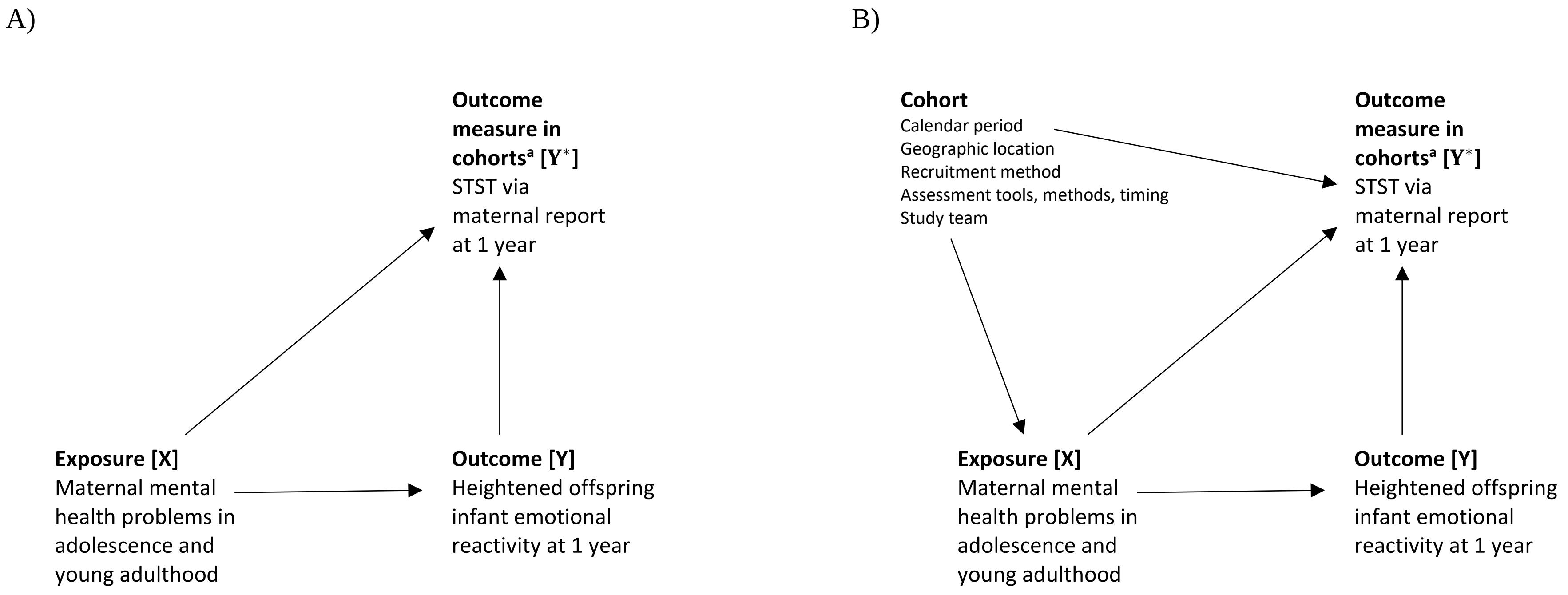

**Figure 3:** Directed Acyclic Graphs (DAGs) depicting examples of: A) measurement bias in a single cohort (within-cohort measurement bias); and B) additional measurement biases in pooled-data analyses of multiple cohorts (across-cohort measurement bias). [a] For the case of $k$ cohorts, there could be up to $k$ different outcome measures $\mathrm{Y}_i^*$, $i = 1, \dots, k$, from which a pooled outcome measure $\mathrm{Y}^{**}$ would be derived. STST: Short Temperament Scale for Toddlers.

**Causal inference in multi-cohort studies using the target trial framework to identify and minimize sources of bias**

**Supplementary Material**

## Description of second case study

O'Connor *et al.* (1) aimed to investigate the extent to which exposure to adversity negatively impacts inflammation in mid to late childhood, where inflammation was proposed as a central mechanism through which exposure to childhood adversity translates to disease risk, in particular cardiovascular disease risk (2, 3).

*Data sources*

Data from two Australian prospective longitudinal cohort studies were utilized.

The Barwon Infant Study (BIS) is a population-derived birth cohort study (*N*=1074 infants) with antenatal recruitment (at approximately 15 weeks of pregnancy) during 2010–2013, conducted in the Barwon region of Victoria, Australia (4). The study was originally designed to explore the early life origins of a range of non-communicable diseases in the modern environment. Participants completed self-reported structured questionnaires as well as clinical and biological measurements at birth and at 1, 6, 9 and 12 months, and at 2 and 4 years, with the most recent primary school (8–10 years) review commencing in 2020–21. Data on inflammatory biomarkers were available for *N* =510 children at the four-year review. Ethical approval for this methodology was obtained from the Barwon Health Human Research Ethics Committee.

The Longitudinal Study of Australian Children (LSAC) is a nationally representative study of two cohorts, including a birth cohort (*N*=5107 infants), aiming to investigate a broad range of aspects of development and wellbeing over the lifecourse, with 9 waves of bi-annual data collection so far, and wave 10 currently underway. In 2003–2004, a multistage cluster sampling design utilising the comprehensive national Medicare database was employed to select a sample that was broadly representative of all Australian children except those living in remote geographic areas (5). In 2015, a comprehensive, one-off physical health and biomarker module, known as the Child Health CheckPoint, was conducted for the birth cohort between waves 6 and 7, when children were 11–12 years of age (6). Approximately half (53%, *N*=1874 families) of the Wave 6 sample participated in the Child Health CheckPoint (7). The study is overseen by the Australian Institute of Family Studies human ethics review board.

*Objectives of multi-cohort design & published findings*

This study aimed to investigate heterogeneity in the causal effect of exposure to adversity on inflammation across the different outcome measurement time-points of mid-childhood (4 years) and late-childhood (11–12 years), for which replication of analyses were performed. Cohort-specific effect estimates were reported and small associations between exposure to adversity and increased inflammation were consistently observed across both cohorts, however, effects were imprecisely estimated.

**Application of the target trial framework**

Supplementary Table 1 outlines a proposed target trial and corresponding emulation strategies for the two cohorts (BIS and LSAC) implicit in the statistical analysis approach described in O'Connor *et al.* (1). The final column of the table identifies potential remaining "within-cohort biases" not addressed within the analysis approach. Note the specific difference in the causal effect definition for which effect heterogeneity is examined, namely the different outcome measurement time-points, is explicitly defined in the follow-up period protocol component of the target trial.

The paper examined multiple definitions of adversity including a binary exposure to each of several different types of adversity, a cumulative count of the types of adversities experienced, and initial timing of exposure to adversity. For simplicity, here we consider a binary indicator of exposure to any type of adversity.

Given this case study focuses primarily on replication of analyses, we consider possible explanations for discrepant findings across cohorts. These could be attributed to discrepant remaining within-cohort biases, detailed in the final column of Supplementary Table 1, chance, or alternatively, may be explained by an actual difference in the causal effect across the two time points at which the inflammation outcome was captured by the studies (mid-childhood at 4 years in BIS vs. late-childhood at 11–12 years in LSAC).

**Supplementary Table 1:** Proposed target trial and emulation strategies implicit in the statistical analysis approach of O'Connor *et al*. (1) for considering the causal effect of exposure to adversity on inflammation in mid- and late-childhood.

| Protocol component | Target trial | Emulation strategies: BIS | Emulation strategies: LSAC | Potential remaining within-cohort bias risks |
|---|---|---|---|---|
| ***A. Eligibility criteria*** | **Target population:**<br>Australian infants at birth in early 2000s | **Analytic sample selection:**<br>BIS participants, who were recruited through pregnant women attending antenatal appointments at approximately 15 weeks during 2010–2013, in Barwon region of Victoria (south-east Australia).<br><br>**Approach to handling missing data and other potential sources of selection bias:**<br>All BIS participants were retained in the sample regardless of missing data via use of multiple imputation. | **Analytic sample selection:**<br>LSAC participants who subsequently participated in the Child Health CheckPoint, a one-off physical health assessment at 11–12 years. LSAC is a cohort of Australian infants aged 0-1 years in 2004 recruited through multi-stage cluster sampling of the comprehensive Medicare database.<br><br>**Approach to handling missing data and other potential sources of selection bias:**<br>All LSAC CheckPoint participants were retained in the sample regardless of missing data via use of multiple imputation. | • Risk of selection bias due to each study's sample selection strategy (e.g., calendar period, geographic location, recruitment procedure) capturing only a subset of the target population (e.g., those able to speak and understand English)<br>• Risk of selection bias due to non-participation:<br>– In BIS, baseline cohort characteristics similar to the Australian population, except a smaller proportion of families from non-English speaking backgrounds<br>– In LSAC, baseline cohort characteristics broadly representative of the Australian population, except a smaller proportion of children living in highly remote geographic areas<br>– In LSAC, participation in Child Health CheckPoint for outcome assessment required presentation at a testing site for venous blood collection; sample was more socially advantaged than the original cohort<br>• Risk of selection bias (in each of BIS and LSAC) due to loss to follow-up/missing data in any analysis variable; mitigated in |
| | | | | both cohorts by use of multiple imputation on all missing data |
| ***B. Treatment strategies*** | **Treatment arms in the trial:**<br>Intervention arm:<br>Experience of adversity during childhood<br>Comparator arm:<br>No experience of adversity during childhood | **Treatment/Exposure measure:**<br>Intervention arm:<br>Exposed to adversity at any measured time point(s) during childhood<br>Comparator arm:<br>Never exposed to adversity during childhood<br><br>Adversity measured as parent-reported presence of any of seven adverse experiences:<br>− Parent legal problems<br>− Parent mental illness<br>− Parent substance abuse<br>− Anger in parenting responses<br>− Separation/divorce<br>− Unsafe neighbourhood<br>− Family member death<br><br>Each adversity measured at least once across the waves (but not at all waves):<br>− W1 (1 month)<br>− W2 (6 months)<br>− W3 (12 months)<br>− W4 (2 years)<br>− W5 (4 years) | **Treatment/Exposure measure:**<br>Intervention arm:<br>Exposed to adversity at any measured time point(s) during childhood<br>Comparator arm:<br>Never exposed to adversity during childhood<br><br>Adversity measured as parent-reported presence of any of seven adverse experiences:<br>− Parent legal problems<br>− Parent mental illness<br>− Parent substance abuse<br>− Anger in parenting responses<br>− Separation/divorce<br>− Unsafe neighbourhood<br>− Family member death<br><br>Each adversity measured at each wave:<br>− W1 (0–1 years)<br>− W2 (2–3 years)<br>− W3 (4–5 years)<br>− W4 (6–7 years)<br>− W5 (8–9 years)<br>− W6 (10–11 years) | • Measurement issues (beyond an imprecisely-defined intervention):<br>− The use of imperfect measures of childhood adversity, e.g., parental mental illness measured in BIS using the Edinburgh Postnatal Depression score>13 (depression likely) and in LSAC using K-6 scale>13 (high psychological distress)<br>− The full range of adversity experienced during childhood not being adequately captured, e.g., racial discrimination<br>− Some adversities measured using proxies, for example, anger in parental responses scale used as a proxy for child maltreatment<br>− Family circumstances and experience of adversity may alter reporting<br>− Adversity indicators sometimes not including the full interval between waves, for example, responses were made in reference to the past 12 months even if waves were >12 months apart, meaning some adverse experiences may not have been captured<br>− In LSAC, a change in scale of measurement for anger in |
| | | | | parental responses (harsh parenting) between waves 2 and 3<br>− In BIS, adversity indicators of unsafe neighbourhood and anger in parenting responses measured at only one wave |
| ***C. Assignment procedures*** | **Randomisation strategy:**<br>Randomisation at recruitment (birth) without blind assignment | **Selection of confounders:**<br>Confounder (self-reported measure)<br>• Child sex<br>• Family socioeconomic position (composite of education and income, dichotomised bottom third vs. higher)<br>• Young maternal age (below or above 23 years)<br>• Indoor smoking (Y/N, same room as baby)<br>• Ethnicity (Anglo/European, Ethnic minority)<br>• BMI (continuous) at 4–5 years<br><br>**Approach to confounding adjustment:**<br>Outcome regression | **Selection of confounders:**<br>Confounder (self-reported measure)<br>• Child sex<br>• Family socioeconomic position (composite of education, occupation and income, dichotomised bottom third vs. higher)<br>• Young maternal age (below or above 23 years)<br>• Indoor smoking (Y/N, any indoor smoking)<br>• Ethnicity (Anglo/European, Ethnic minority)<br>• BMI (continuous) at 4–5 years<br><br>**Approach to confounding adjustment:**<br>Outcome regression | • Risk of residual confounding bias due to unmeasured confounding<br>• Risk of residual confounding bias due to measurement error:<br>− The use of proxies for confounders in the absence of more direct measures, e.g., a composite variable of education, occupation and income for socioeconomic position, a composite of language and country of birth for ethnicity, indoor smoking measured using Y/N same room as baby in BIS vs. Y/N any indoor smoking in LSAC<br>− Inaccurate reporting of confounders that reveal sensitive information such as income, smoking history |
| ***D. Follow-up period*** | **Start and end times:**<br>Start: At birth<br><br>*Difference of interest in effect heterogeneity analysis:*<br>Endpoint 1: Mid-childhood (4 years) | **Timing of measures:**<br>Start: Wave 0, pregnancy<br>Ends: Wave 5, 4 years | **Timing of measures:**<br>Start: Wave 1, 0–1 years<br>Ends: Wave 6.5 (Child Health CheckPoint), 11–12 years | • Measurement issues due to the exposure not being measured continuously over all childhood nor in the same way for all participants<br>• Risk of measurement bias due to the outcome not being measured at exactly the specified endpoint |

| | | | | | |
|---|---|---|---|---|---|
| | | Endpoint 2: Late-childhood (11–12 years) | | | (4 years for BIS, 11–12 years for LSAC)<br>Note: The difference in the outcome measurement endpoints is the key factor of interest in the research question, therefore it is not a bias per se, but the source of difference to be assessed |
| ***E.*** | ***Outcome*** | **Outcome measure:**<br>Inflammatory markers (continuous, µg/ml):<br>– hsCRP<br>– GlycA | **Outcome measure:**<br>Inflammatory markers (continuous, µg/ml):<br>– hsCRP<br>– GlycA | **Outcome measure:**<br>Inflammatory markers (continuous, µg/ml):<br>– hsCRP<br>– GlycA | |
| ***F.*** | ***Causal contrasts of interest and causal effect measure*** | Percentage difference in mean inflammatory marker levels between intervention and comparator arms in the target population | | | |